\documentclass[twocolumn,showpacs,floats,floatfix,superscriptaddress,aps,pra]{revtex4}
\usepackage{amsfonts}
\usepackage{amssymb}
\usepackage{amsmath}
\usepackage{calc,graphicx,color,bm}

\def\be{ \begin{equation} }
\def\ee{ \end{equation} }
\def\bea{ \begin{eqnarray} }
\def\eea{ \end{eqnarray} }
\def\bse{ \begin{subequations} }
\def\ese{ \end{subequations} }
\def\sech{\,\text{sech}\,}

\def\i{\,\text{i}}
\def\e{\,\text{e}}

\def\to{\rightarrow}

\def\half{\tfrac12}

\newcommand{\ket}[1]{\vert #1\rangle}

\def\t{r}
\def\U{\mathbf{U}}
\def\F{\mathbf{F}}
\def\H{\mathbf{H}}

\def\adphase{\alpha}
\def\adstate{\chi}
\def\adenergy{\varepsilon}

\begin{document}

\author{Boyan T. Torosov}
\affiliation{Department of Physics, Sofia University, James Bourchier 5 blvd, 1164 Sofia, Bulgaria}
\author{Nikolay V. Vitanov}
\affiliation{Department of Physics, Sofia University, James Bourchier 5 blvd, 1164 Sofia, Bulgaria}
\affiliation{Institute of Solid State Physics, Bulgarian Academy of Sciences, Tsarigradsko chauss\'{e}e 72, 1784 Sofia, Bulgaria}
\title{Design of quantum Fourier transforms and quantum algorithms by using circulant Hamiltonians}
\date{\today }

\begin{abstract}
{We propose a technique for design of quantum Fourier transforms, and ensuing quantum algorithms, in a single interaction step
by engineered Hamiltonians of circulant symmetry. The method uses adiabatic evolution and is robust against fluctuations of the
interaction parameters as long as the Hamiltonian retains a circulant symmetry.}
\end{abstract}


\maketitle

\section{Introduction}

Quantum information processing is built upon sequences of special unitary transformations. %
One of the most important of these is the quantum (discrete) Fourier transform (QFT), which is a key ingredient of many
quantum algorithms \cite{Nielsen&Chuang,Josza98,Bowden02}, including Shor's factorization \cite{Shor},
 the algorithms of Deutsch \cite{Deutsch} and Simon \cite{Simon}, order finding, discrete logarithms, quantum phase estimation, etc. \cite{Nielsen&Chuang}.

Traditionally, QFT on $\t$ qubits is implemented by a quantum circuit consisting of $O(\t)$ Hadamard gates and $O(\t^2)$ controlled-phase gates \cite{Coppersmith}.
Experimental demonstrations include
 synthesis of 3-qubit QFT in nuclear magnetic resonance (NMR) systems \cite{Weinstein01},
 order finding with NMR \cite{Vandersypen00},
 phase estimation with NMR \cite{Lee02},
 Shor's factorization in NMR \cite{Vandersypen01},
 in ion traps \cite{Chiaverini05},
 and using a ``compiled version'' of Shor's algorithm with photonic qubits \cite{Lanyon07,Lu07}.
Further theoretical proposals for implementations of QFT include %
 atoms in cavity QED \cite{Scully&Zubairy}, %
 entangled multilevel atoms \cite{Muthukrishnan02}, %
 trapped ions with Householder reflections \cite{Ivanov08}, %
 linear optics \cite{Barak07} with Cooley-Tukey's algorithm \cite{Cooley-Tukey}, %
 waveguide arrays \cite{Akis01}, etc.

The largest numbers factorized experimentally by Shor's algorithm hitherto are 15 \cite{Vandersypen01,Chiaverini05} and 21 \cite{Peng08}.
The primary obstacle for demonstration of Shor's factorization for larger numbers is the large number of one- and two-qubit gates required.
A ``general-purpose'' Shor's algorithm for an L-bit number demands $L^3$ gates and $5L+1$ qubits \cite{Beckman};
 an implementation using a linear ion trap would require about $396L^3$ laser pulses \cite{Beckman}.
``Special-purpose'' algorithms that exploit special properties of the input number are much faster:
 the number 15 can be factored with 6 qubits and 38 pulses only \cite{Beckman}.

Another practical difficulty of the QFT algorithm is the use of two-qubit control-phase gates, which, for large number of qubits, involve very small phases.
To this end, an ``approximate'' QFT has been proposed \cite{Coppersmith,Barenco96,Cheung04}, in which the phase shift gates requiring highest precision are omitted.

Griffiths and Niu proposed a ``semiclassical'' QFT, wherein the costly two-qubit gates
 are replaced by serial single-qubit rotations supplemented with classical measurements \cite{Griffiths96}. Such a semiclassical QFT has been demonstrated recently with three trapped ions \cite{Chiaverini05}.

In the present work, we propose to construct QFT by a novel approach which uses a special class of Hamiltonians, having a circulant symmetry. {Such}
Hamiltonians have the advantage that their eigenvectors are the columns of the QFT {(hence the latter
diagonalizes the Hamiltonian)}, and they \emph{do not depend} on the particular elements of the Hamiltonian, as far as the
{circulant} symmetry is conserved. This important feature allows one to construct QFT in a \emph{single interaction
step}; it also makes this techniques 
 robust against variations in the interaction parameters.
The present paper uses a similar approach as Unanyan \emph{et al.} \cite{Unanyan}, who proposed to use circulant Hamiltonians in
order to create coherent superpositions of states.

\section{Background}

\paragraph*{Quantum Fourier transform.} The $N$-dimensional QFT is defined with its action on an orthonormal basis $\ket{0},\ket{1},\ldots ,\ket{N-1}$: 
\be
\label{Fourier} \F^N\ket{n}=\frac{1}{\sqrt{N}}\sum_{k=0}^{N-1}\e^{2\pi\i nk/N}\ket{k} .
\ee%
{It} transforms a single state into an equal superposition of states with specific phase factors. The inverse QFT is%
\be\label{Inverse Fourier}
(\F^N)^{-1}\ket{n}=\frac{1}{\sqrt{N}}\sum_{k=0}^{N-1}\e^{-2\pi\i nk/N}\ket{k} .
\ee%
In a matrix form QFT is a square matrix with elements%
\be\label{Fourier matrix} \F_{kn}^N=\frac{1}{\sqrt{N}}\e^{2\pi \i kn/N} . \ee%

\paragraph*{Circulant matrix.}
An $N\times N$ matrix $C$ of the form
\be\label{circulant matrix}
C= \left[
\begin{array}{ccccc}
c_0 & c_{N-1} & c_{N-2} & \cdots & c_1 \\
c_1 & c_0 & c_{N-1} & \cdots & c_2 \\
c_2 & c_1 & c_0 & \cdots & c_3 \\
\vdots & \vdots & \vdots & \ddots & \vdots \\
c_{N-1} & c_{N-2} & c_{N-3} & \cdots & c_0
\end{array}%
\right] \ee%
is called a circulant matrix. It is a special case of a Toeplitz matrix \cite{Toeplitz} and it is completely defined by its
first vector-column (or row). The other columns (rows) are just cyclic permutations of it. The circulant matrices have some very
interesting properties. The most important{ one in the present context} is that the eigenvectors of a circulant matrix of a
given size are the vector-columns of the discrete Fourier transform
{\eqref{Fourier matrix}} of the same size; {hence they do not depend on the elements of the circulant matrix.} %
The eigenvalues $\lambda_n$ of the circulant matrix, { though, are phased sums of the matrix elements:}%
\be\label{eigenvalues} \lambda_n =\sum_{k=0}^{N-1}c_k \exp \left( -\i 2\pi kn/N \right) . \ee

\section{Design of the Hamiltonian}

In order to {synthesize} QFT, we use a special time-dependent Hamiltonian of the form \cite{Unanyan} %
\be\label{Hamiltonian} \H (t)=f(t)\H_0+g(t)\H_1 , \ee%
where $f(t)$ and $g(t)$ are (generally pulse-shaped) real-valued functions, such that $f(t)$ \emph{precedes} $g(t)$ in time, i.e. 
\be
0 \overset{-\infty \leftarrow t}{\longleftarrow} \frac{g(t)}{f(t)} \overset{t \rightarrow \infty}{\longrightarrow} \infty. %
\ee%
For instance, we can take %
\bse\label{fg} \bea\label{f}
&& f(t) = [1-\tanh(t/T)]/2 ,\\
&& g(t) = [1+\tanh(t/T)]/2 . \label{g}
\eea\ese %
Therefore, the Hamiltonian \eqref{Hamiltonian} has the asymptotics%
\be\label{Hamiltonian asymptotics}%
\H_0 \overset{-\infty \leftarrow t}{\longleftarrow} \H(t)\overset{t \rightarrow \infty}{\longrightarrow} \H_1 . %
\ee%

We demand $\H_0$ to be a diagonal matrix in which the energies of all states (the diagonal elements) are non-degenerate%
\be \H_0 = \text{diag}(E_1,E_2,\ldots,E_N). \ee %
For $\H_1$ we choose a circulant matrix, with the condition that the eigenvalues should be well separated from each other.
Because the Hamiltonian has to be Hermitian, $\H_1$ is not a most general circulant matrix, but a Hermitian circulant matrix.

Because the Hamiltonian {\eqref{Hamiltonian}} at $t\to\infty$ has a circulant symmetry, its eigenvectors are the vector-columns
of QFT. However, {each} eigenvector {$\ket{n}$} may have an adiabatic phase {factor $\e^{\i\adphase_n}$}, {acquired in the end
of the evolution, which may be different for each $\ket{n}$}. This means that for such a Hamiltonian \eqref{Hamiltonian},
adiabatic evolution will perform the QFT \eqref{Fourier} ({possibly} after {renumbering}
of the basis states), but with some {additional phases $\adphase_n$,} 
\be\label{Fourier2} \F^N\ket{n}=\frac{1}{\sqrt{N}}\e^{\i\adphase_{n}}\sum_{k=0}^{N-1}\e^{2\pi\i { n}k/N}\ket{k} . \ee%
The phases $\adphase_n$ are just integrals over the quasienergies, {as follows from} the adiabatic
theorem \cite{Messiah}. The inverse Fourier transform would be %
\be\label{Inverse Fourier 2} (\F^N)^{-1}\ket{n} = \frac{1}{\sqrt{N}}\sum_{k=0}^{N-1}\e^{-\i\adphase_k}\e^{-2\pi\i {n}k/N}\ket{k}, \ee%
{and} it can be accomplished by adiabatic evolution with the Hamiltonian %
\be\label{Hamiltonian2} \H(t)=g(t)\H_0+f(t)\H_1 . \ee %

For example, for $N=4$ {we can have} %
\bse \bea\label{H0}%
&&\H_0 =\text{diag} (-E,-E/3,E/3,E) \\ \label{H1}%
&& \H_1 = \left[\begin{array}{cccc}
0 & V & 0 &  V^{\ast} \\
V^{\ast} & 0 & V & 0  \\
0 & V^{\ast} & 0 & V   \\
V & 0 & V^{\ast} & 0
\end{array}\right] .
\eea
\ese%
For laser-driven atomic and molecular transitions, the interaction $V$ is given by the Rabi frequency $\Omega$:
$V=\frac12\hbar\Omega$. {Insofar} as the eigenvalues of the circulant matrix are given by Eq. \eqref{eigenvalues}, {one has} to
choose the interaction {energy} $V$ in such {a} way that the eigenenergies have well separated values. Another requirement for adiabatic evolution is that the functions $f(t)$ and $g(t)$ change sufficiently slowly, so that the
nonadiabatic coupling $\langle\dot\adstate_m(t) \mid \adstate_n(t) \rangle$ between each pair of adiabatic states
$\ket{\adstate_m(t)}$ and $\ket{\adstate_n(t)}$ remains negligibly small compared to the separation of the eigenenergies
$\adenergy_m(t)$ and $\adenergy_n(t)$,
\be%
|\adenergy_m(t)-\adenergy_n(t)| \gg |\langle \dot\adstate_m(t) \mid \adstate_n(t) \rangle | \sim \frac 1 T,
\ee%
where 
 $T$ is the interaction duration.

\begin{figure}[tb]
\includegraphics[width=70mm]{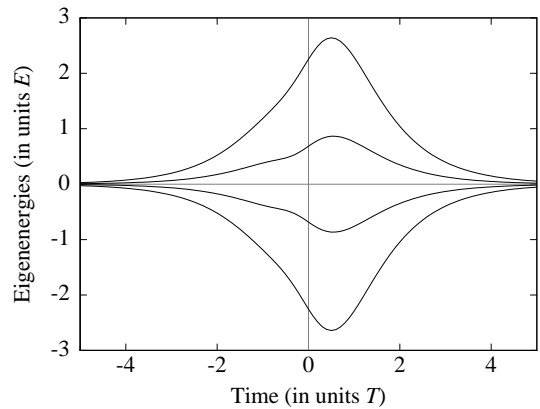}
\caption{Eigenvalues of the Hamiltonian \eqref{Hamiltonian} as a function of time for $\H_0$ and $\H_1$ given by Eqs. \eqref{H0}
and \eqref{H1}, with $V=E(1+\i/3)$, whereas $g(t)$ and $f(t)$ are given by Eqs. \eqref{f2} and \eqref{g2}.} \label{Eigen4}
\end{figure}

In the numeric examples we use a hyperbolic-secant mask for the functions $f(t)$ and $g(t)$:
\bse\label{fg2} \bea\label{f2}
&& f(t) = \sech(t/\tau) [1-\tanh(t/T)] ,\\
&& g(t) = \sech(t/\tau) [1+\tanh(t/T)] . \label{g2}
\eea\ese
These factors are chosen for implementation feasibility; they do not change the (all-important) asymptotic behaviour of the eigenstates $\ket{\chi_n(t)}$.
Figure \ref{Eigen4} shows the evolution of the eigenvalues of the Hamiltonian \eqref{Hamiltonian} for $V=E(1+\i/3)$. %
For this choice of $V$ the eigenenergies are non-degenerate (except at infinite times, which is irrelevant because there
is no interaction) and the adiabatic evolution is enabled.

\section{Quantum phase estimation}

We shall show now that the QFT propagator \eqref{Fourier2}, which results from the Hamiltonian \eqref{Hamiltonian}, can be used
to realize quantum algorithms, despite the presence of the adiabatic phase factors $\e^{\i\adphase_{ n}}$. We consider the
quantum phase estimation algorithm \cite{Nielsen&Chuang}, which is the key for many other algorithms, such as Shor's
factorization. We briefly summarize here the essence of this algorithm.

Let us consider a unitary operator $\U$, which has an eigenvector $\ket{u}$ and a corresponding eigenvalue $\exp (2\pi\i\phi)$,
where $\phi\in[0,1)$. We assume that we are able to prepare state $\ket{u}$ and to perform the controlled-$\U^{2^j}$ operation, for
non-negative integer $j$. The goal of the algorithm is to estimate $\phi$. To this end, the algorithm uses two registers. The
first register contains $\t$ qubits initially in state $\ket{0}$ and the second one starts in state $\ket{u}$, containing as
many qubits as needed to store $\ket{u}$.

The procedure starts with the application of a Hadamard transform \cite{Nielsen&Chuang} to the first register, followed by the
application of controlled-$\U$ operations on the second register, with $\U$ raised to successive powers of two.
The final state of the first register is %
\bea\notag
{2^{-\t/2}} (\ket{0}+\e^{2\pi \i 2^{\t-1}\phi}\ket{1}){_1}%
 (\ket{0}+\e^{2\pi \i 2^{\t-2}\phi}\ket{1}){_2} \cdots \notag\\
 \times (\ket{0}+\e^{2\pi \i 2^0\phi}\ket{1}){_r}  \label{state 1st reg}
= {2^{-\t/2}} \sum_{k=0}^{2^\t-1}\e^{2\pi \i k \phi }\ket{k},
\eea%
and the second register stays in state $\ket{u}$. Now let us suppose that $\phi$ can be expressed using a $\t$-bit expansion%
\be\label{t-bit} \phi = 0.\phi_1\phi_2\ldots\phi_\t  = \frac{\phi_1}{2}+\frac{\phi_2}{4}+\cdots +\frac{\phi_\t}{2^\t} , \ee%
where $0.\phi_1\phi_2\ldots\phi_\t$ represents a binary fraction. %
{Then} state \eqref{state 1st reg} can be written as %
\bea
2^{-\t/2} (\ket{0}+\e^{2\pi \i 0.\phi_\t}\ket{1}){_1}%
 (\ket{0}+\e^{2\pi \i 0.\phi_{\t-1}\phi_\t}\ket{1}){_2} \cdots \notag\\
 \times (\ket{0}+\e^{2\pi \i 0.\phi_1\phi_2\cdots\phi_\t}\ket{1})_r .
\eea%
Finally, we apply the inverse QFT in order to obtain the product state $\ket{\phi_1\cdots\phi_\t}$.
{In our case we apply} the {phased} inverse QFT \eqref{Fourier2} and {find} %
\be\label{phase} \e^{-\i\adphase(\phi)}\ket{\phi_1\cdots\phi_\t} , \ee%
where $\adphase(\phi)$ is an adiabatic phase that depends on $\phi$. Since this global phase $\adphase$ has no physical meaning,
a measurement in the computational basis would give us exactly $\phi$. We note that if $\phi$ cannot be written as a $\t$ bit
expansion \eqref{t-bit}, this procedure can still produce a good approximation to $\phi$ with high probability
\cite{Nielsen&Chuang}.

\begin{figure}[tb]
\includegraphics[width=70mm]{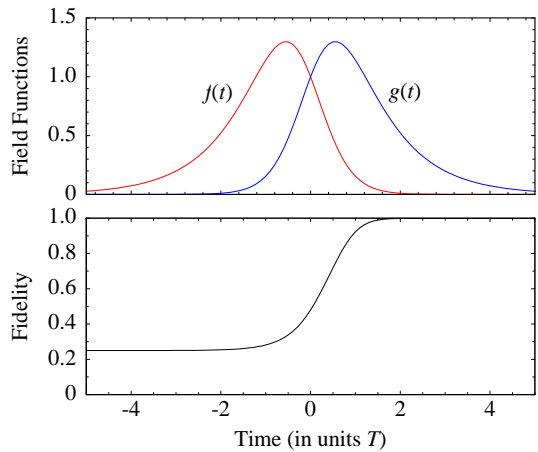}
\caption{Upper frame: Field functions $f(t)$ and $g(t)$, Eqs.~\eqref{f2} and \eqref{g2}. Lower frame: Fidelity of phase estimation
during the inverse Fourier transform. The parameters are: $\phi=0.75$, $\tau=T$, $V=E(1+\i/3)$, $E=10/T$.} \label{Fidelity}
\end{figure}

In Fig. \ref{Fidelity} we plot the probability of state \eqref{phase} during the inverse Fourier transformation \eqref{Inverse
Fourier 2}. This probability is evaluated by solving numerically the Schr\"{o}dinger equation for the Hamiltonian
\eqref{Hamiltonian2} and is used as a measure of the fidelity. {The figure shows} that when the phase $\phi$ has an exact
expansion as a binary fraction, the final probability tends to unity.

\section{Implementations}

\begin{figure}[tb]
\includegraphics[width=80mm]{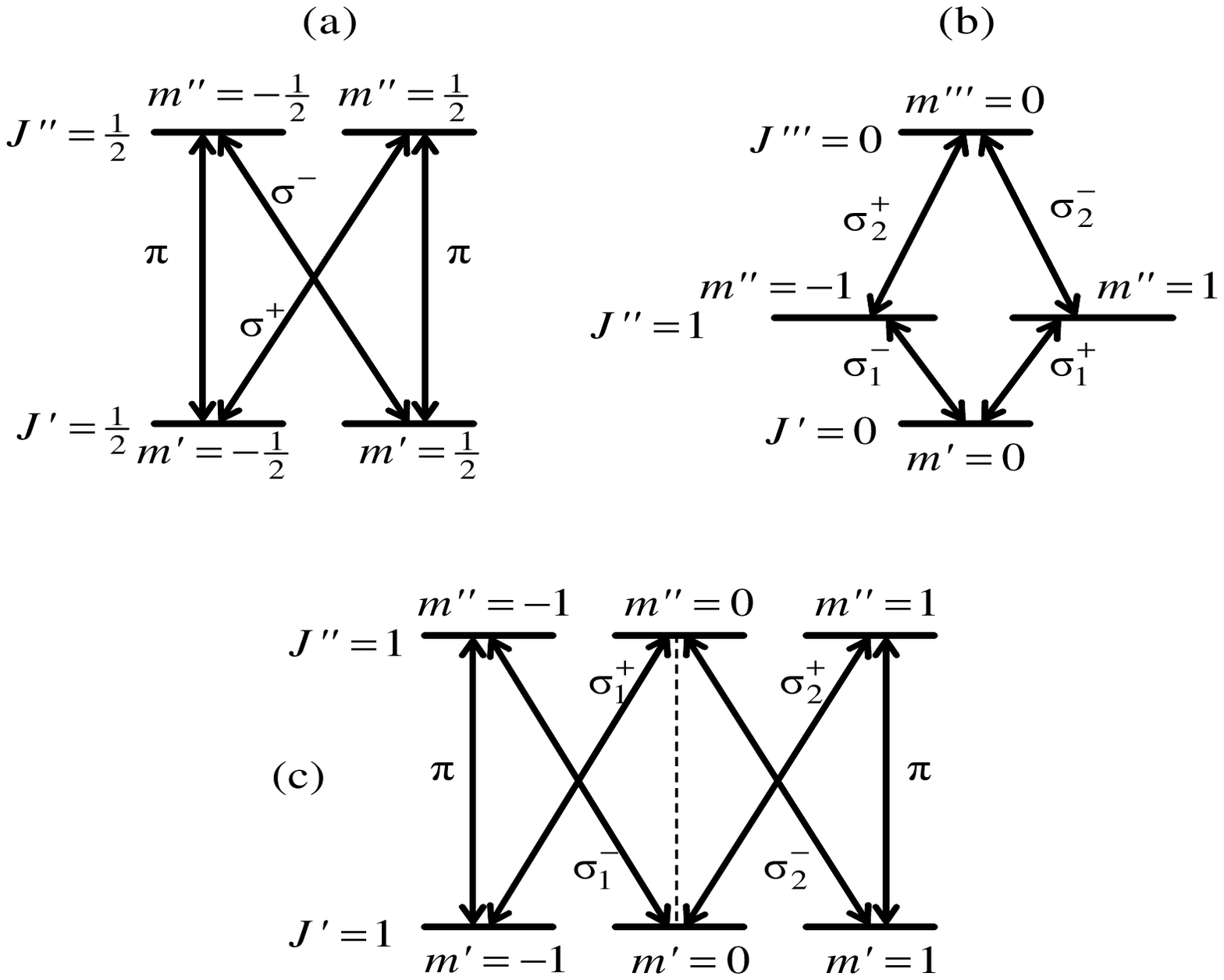}
\caption{Systems, which can be used in order to realize a circulant Hamiltonian: (a) $J^\prime=\frac12\leftrightarrow J^{\prime\prime}=\frac12$
system, (b) $J^\prime=0\leftrightarrow J^{\prime\prime}=1\leftrightarrow J^{\prime\prime\prime}=0$ system, and (c)
$J^\prime=1\leftrightarrow J^{\prime\prime}=1$ system, where the $m^\prime=0\leftrightarrow m^{\prime\prime}=0$ transition is dipole forbidden.} \label{J}
\end{figure}

In this section we discuss a few simple systems, which can be used for implementing the Hamiltonian {\eqref{Hamiltonian}}. %

\paragraph*{$J^\prime=\frac12\leftrightarrow J^{\prime\prime}=\frac12$ system.}

As a first example we consider the system formed of the magnetic sublevels in a $J^\prime=\frac12\leftrightarrow
J^{\prime\prime}=\frac12$ transition shown in Fig. \ref{J}(a),
 where $J$ is the total angular momentum of each level.
We apply two linearly polarized fields (the two polarization directions being perpendicular), the second one seen as a
superposition of two circularly polarized fields ($\sigma_+$ and $\sigma_-$). %
By ordering the magnetic sublevels in the sequence $\ket{m^\prime=-\frac12}$, $\ket{m^{\prime\prime}=\frac12}$,
$\ket{m^\prime=\frac12}$, $\ket{m^{\prime\prime}=-\frac12}$, and 
by suitably tuning the strengths and the relative phase of the two independent fields, we can adjust the interaction elements of
the Hamiltonian and produce the desired circulant form \eqref{H1}. %
We note that because of the different signs of some of the Clebsch-Gordan coefficients, one should redefine one of the
probability amplitudes by changing its sign.

As we want to implement the full Hamiltonian \eqref{Hamiltonian}, we also need to realize its first part $\H_0$. %
It is especially important to remove the degeneracies between the magnetic sublevels. %
This can be accomplished by using a static magnetic field, which induces $m$-dependent Zeeman shifts, and a far-off resonant
laser pulse, which will cause Stark shifts. Let the energy splitting due to Zeeman shift {be} $E_{\text{Z}}$ (the same for both
ground and excited levels). The Stark shifts are generally different for the two levels: $E_{\text{g,S}}$ and $E_{\text{e,S}}$,
where `g' and `e' stand, respectively, for ground and excited. Hence, in order to realize the Hamiltonian \eqref{H0}, we need to
solve the following algebraic system
\bse\bea %
-\half E_{\text{Z}} + E_{g, \text{S}}&=&-E, \\
\half E_{\text{Z}} + E_{g, \text{S}}&=&-E/3, \\
-\half E_{\text{Z}} + E_{e, \text{S}}&=&E/3, \\
\half E_{\text{Z}} + E_{e, \text{S}}&=&{E} , %
\eea\ese%
which gives $E_{\text{Z}}=E_{\text{e,S}} = -E_{\text{g,S}} = \frac23E$. {Moreover}, because the energies of {$\mathbf{H}_0$}
need not be exactly evenly separated, our method is robust against
fluctuations in the field parameters. %
Making a reference to Eq.~\eqref{Hamiltonian} and Fig.~\ref{Fidelity}, we conclude that the Stark and Zeeman fields, with the
time dependence $f(t)$, have to be applied before the polarized laser fields, with time dependence $g(t)$.

\paragraph*{$J^\prime=0 \leftrightarrow J^{\prime\prime}=1 \leftrightarrow J^{\prime\prime\prime}=0$ system.}

Another system with $N=4$ is the diamond system depicted in Fig. \ref{J}(b). Here again two linearly polarized {laser} fields
are needed, but {now} they have parallel polarization directions. One advantage of this system is that only magnetic fields are
sufficient to realize the first part of the Hamiltonian. The disadvantage is that the two independent fields generally come from
two different lasers, because of the different frequencies of the transitions.

\paragraph*{$J^\prime=1\leftrightarrow J^{\prime\prime}=1$ system.}

The $J^\prime=1\leftrightarrow J^{\prime\prime}=1$ system, depicted in Fig. \ref{J}(c), contains six coupled $m$ sublevels. In
this system the circulant symmetry occurs because $m^\prime=0 \leftrightarrow m^{\prime\prime}=0 $ is a dipole forbidden transition. %
By ordering the magnetic sublevels in the sequence %
$\ket{m^\prime=-1}$, $\ket{m^{\prime\prime}=0}$, $\ket{m^\prime=1}$, $\ket{m^{\prime\prime}=1}$,
$\ket{m^\prime=0}$, $\ket{m^{\prime\prime}=-1}$ %
we obtain a Hamiltonian of the type %
\be \H_1 = \frac \hbar 2 \left[ \begin{array}{cccccc}
0 & -\Omega_1 & 0 & 0 & 0 & -\Omega_2 \\
-\Omega_1^{\ast} & 0 & \Omega_1^{\ast} & 0 & 0 & 0 \\
0 & \Omega_1 & 0 & \Omega_2 & 0 & 0  \\
0 & 0 & \Omega_2^{\ast} & 0 & -\Omega_1^{\ast} & 0 \\
0 & 0 & 0 & -\Omega_1 & 0 & \Omega_1 \\
-\Omega_2^{\ast} & 0 & 0 & 0 & \Omega_1^{\ast} & 0
\end{array} \right] ,
\ee%
where $\Omega_1$ and $\Omega_2$ are the Rabi frequencies between, respectively, states with different $m$ and states with the same $m$. %
Each Clebsch-Gordan {coefficient} is incorporated in the respective Rabi frequency. %
The Rabi frequencies are complex (needed to avoid eigenvalue degeneracies), with a phase difference between the left and right circularly polarized components. %
After a phase transformation of the amplitudes, $c_n\rightarrow\e^{\i\beta_n}c_n $, with suitably chosen phase factors
$\beta_n$, we can make the Hamiltonian take the form of a circulant matrix. The selection of the phases $\beta_n$ amounts to
solving a simple linear algebraic system.

The first part of the Hamiltonian {$\H_0$} {can be realized with auxiliary magnetic and electric fields, as for} the
$J^\prime=\frac12\leftrightarrow J^{\prime\prime}=\frac12$ system.

\section{Conclusions}

The intrinsic symmetry of circulant matrices allows one to design Hamiltonians that can produce a discrete Fourier transform on
a set of quantum states in a natural manner and in a single step, without the need to apply a large number of consecutive
quantum gates. %
The designed Hamiltonian has different asymptotics: it is a nondegenerate diagonal matrix in the beginning and a circulant
matrix in the end (or vice versa); the time dependence that connects the two should be sufficiently slow in order to enable
adiabatic evolution. The resulting unitary transformation, which this Hamiltonian produces, differs from the standard QFT by
additional (adiabatic) phase factors in the matrix columns; we show, however, that one can still construct the quantum phase
estimation algorithm,
which is an essential subroutine in many quantum algorithms. %
We have presented examples of simple atomic systems, the Hamiltonians of which can be tailored to obtain circulant symmetry. %
The construction of large-scale systems with circulant symmetry requires the design of a \emph{closed-loop} linkage pattern; %
 for instance, a chain of nearest-neighbor interactions supplemented with a (direct or effective) interaction between the two ends of the chain.

\acknowledgments

This work has been supported by the European Commission projects EMALI and FASTQUAST and the Bulgarian NSF grants VU-F-205/06,
VU-I-301/07, D002-90/08, and IRC-CoSiM.


\end{document}